# Gap-dependent quasiparticle dynamics and coherent acoustic phonons in CaFe$_2$As$_2$ across spin density wave phase transition

Sunil Kumar[1,§], Luminita Harnagea[2], Sabine Wurmehl[2], Bernd Buchner[2], and A. K. Sood[1,*]

[1]*Department of Physics, Indian Institute of Science, Bangalore 560012, India*
[2]*Leibniz-Institute for Solid State and materials Research, D-01171 Dresden, Germany*

We report ultrafast quasiparticle (QP) dynamics and coherent acoustic phonons in undoped CaFe$_2$As$_2$ iron pnictide single crystals exhibiting spin-density wave (SDW) and concurrent structural phase transition at temperature T$_{SDW}$ ~ 165 K using femtosecond time-resolved pump-probe spectroscopy. The contributions in transient differential reflectivity arising from exponentially decaying QP relaxation and oscillatory coherent acoustic phonon mode show large variations in the vicinity of T$_{SDW}$. From the temperature-dependence of the QP recombination dynamics in the SDW phase, we evaluate a BCS-like temperature dependent charge gap with its zero-temperature value of ~(1.6±0.2)k$_B$T$_{SDW}$, whereas, much above T$_{SDW}$, an electron-phonon coupling constant of ~0.13 has been estimated from the linear temperature-dependence of the QP relaxation time. The long-wavelength coherent acoustic phonons with typical time-period of ~100 ps have been analyzed in the light of propagating strain pulse model providing important results for the optical constants, sounds velocity and the elastic modulus of the crystal in the whole temperature range of 3 K to 300 K.

## 1. Introduction

Since the recent discovery of high temperature superconductivity in iron pnictides[1] a lot of interest has been generated, both theoretically and experimentally, to understand the magnitude, nature and symmetry of energy gaps in the SDW and superconducting (SC) states of these compounds.[2-6] Parent compounds of the iron pnictide family show a spin-ordered phase below the SDW phase transition temperature T$_{SDW}$ and superconductivity evolves either by electron or hole-doping at a lower temperature. Therefore, the parent compounds are being studied using various experimental methods to understand the role of spin fluctuations, spin-phonon and electron-phonon interactions. The picture on the nature and presence or absence of charge gaps opening at or below the magneto-structural transition in the parent compound so far remains inconclusive. On one side, the temperature-dependent resistivity indicates metallic behavior in the SDW phase.[7,8] Similarly, strong orbital-dependent reconstruction of the Fermi surface across the magneto-structural transition and metallic SDW state[9] or absence of charge gap[10,11] was derived from angle resolved photoemission spectroscopy (ARPES). On the other hand, in many infrared absorption[3,4] and ARPES[6] studies two charge gaps with 2Δ$_0$/k$_B$T$_{SDW}$ ranging between 3.5 to 11 have been reported. In the last couple of years, a few ultrafast time-resolved spectroscopic studies have been reported on the nature of charge gaps in the parent SDW iron pnictides: BaFe$_2$As$_2$,[12] SmFeAsO[13] and SrFe$_2$As$_2$.[14] The transient reflectivity measurements have shown a bi-exponential dynamics where a fast component could be described by quasiparticle recombination dynamics across a charge gap within the phonon bottleneck model giving band gap value of 2Δ$_0$/k$_B$T$_{SDW}$ ~ 5 and 7.2.[13,14]

Our present study addresses the temperature and fluence-dependent QP dynamics and coherent acoustic phonons in parent CaFe$_2$As$_2$ single crystals investigated by measuring transient differential reflectivity using femtosecond pump-probe spectroscopy which has not been carried out hitherto. The QP dynamics involves three distinct exponentially decaying relaxation channels with decay times varying from sub-picosecond to hundreds of picoseconds where drastic changes occur in the amplitudes and corresponding decay-times at T$_{SDW}$. Such a behavior is clearly an indication of charge gap opening in the spin density wave phase. By analyzing the temperature-dependence of the fast relaxation component using phonon bottleneck model[15,16] we derive the zero-temperature gap value of 2Δ$_0$ ~ (1.6±0.2)k$_B$T$_{SDW}$. In the high temperature normal metallic phase of the sample, linear temperature-dependence of the QP relaxation time has been used to estimate the coupling strength of electrons with the optical phonons to be ~0.13. Further, our experimental results clearly show detection of very low-frequency (GHz) coherent oscillations superimposed on slowly exponentially decaying background. These are attributed to coherent excitation of an acoustic phonon mode launched by laser induced electronic and/or thermal stress at the sample surface. Analysis of the observed mode using strain pulse propagation model in the ultrafast ultrasonics[17] yields temperature dependence of the optical constants, sound velocity and hence the elastic modulus across the SDW phase transition. Moreover, from the phonon-amplitude peaking at a temperature higher than T$_{SDW}$, we infer strong magneto-elastic coupling between spin-fluctuations in the normal paramagnetic phase and the lattice, a behavior similar to that observed in multiferroic manganites.[18,19]

## 2. Experimental details

Single crystals of CaFe$_2$As$_2$ were grown by high temperature solution growth technique using Sn flux and characterized as reported earliar.[8] Spin density wave phase transition along with a concurrent structural (S) transition from high symmetry tetragonal phase to orthorhombic symmetry at low temperatures were established to occur at $T_{SDW}$ ($T_S$) ~ 165 K. The crystals with c-axis perpendicular to the crystal surface were cleaved into platelet samples with thickness ~0.3 mm. Degenerate pump-probe experiments in a non-collinear geometry were carried out at 790 nm using 45 fs laser pulses at 1 kHz repetition rate taken from a regenerative amplifier (Spitfire, Spectra Physics). We have used unfocused pump and probe beams with the laser spot size on the sample of ~0.028 cm$^2$. The pump-induced changes (ΔR) in the probe reflectivity R were detected in a usual lock-in detection scheme as a function of time-delay between the pump and probe pulses configured for cross-polarization. For pump-fluence dependent studies we used neutral density filters in the pump path. Continuous flow liquid-helium optical-cryostat was used to vary the sample temperature from 3.4 K to room temperature for temperature-dependent measurements.

## 3. Results

Experimental time-resolved differential reflectivity data (ΔR/R) at various sample temperatures using pump-fluence of ~85 μJ/cm$^2$ are presented in Fig. 1. Overall, the transients can be consistently fitted using a combination of three (k = 1 to 3) exponentially decaying contributions with amplitudes $A_k$ and time-constants $\tau_k$ and a single damped oscillatory part:

$$\frac{\Delta R}{R} = (1 - e^{-t/\tau_R}) \left[ \sum_k A_k e^{-t/\tau_k} + B e^{-t/\tau_p} \cos(2\pi \nu_p t + \phi) \right] \quad (1)$$

Here, $\tau_R$ ~ 100 fs is the rise time of the transients found to be independent of temperature and fluence. Solid lines in Fig. 1 are the fits. The exponentially decaying contributions are attributed to QP recombination/relaxation components and the slow coherent oscillations are due to longitudinal acoustic phonon mode having phonon amplitude B, dephasing time $\tau_p$, frequency $\nu_p$ and initial phase $\phi$. To make the fitting procedure better clear, in Fig. 2 we plot the experimental data (open circles) and the fit (dark line) along with individual components (thin lines) for 3.4 K using Eq. (1). From such an analysis, we describe in detail the temperature- and fluence-dependences of the QP dynamics in Section 3.1 and the coherent acoustic phonon dynamics in Section 3.2.

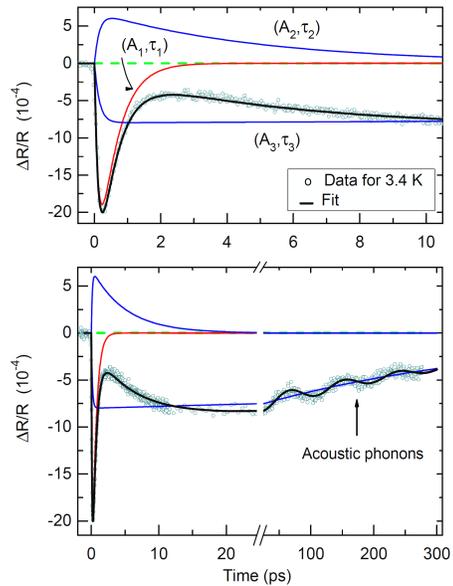

Fig. 2. (Color online) Experimental data (open circles) and fit (solid line) along with individual electronic components (thin lines) in Eq. (1) shown for 3.4 K presented for clarity of the data fitting procedure.

We have used strain pulse propagation model in Section 3.2 to derive temperature dependence of the sound velocity $u_s$ and hence the elastic modulus Y from the observed phonon frequency $\nu_p$ and dephasing time $\tau_p$. This, along with the temperature independent reflectivity value that is measured by focusing the femtosecond laser beam (low average power) and collecting all the reflected light for near-normal incidence angle allows us to estimate the refractive index n and the extinction coefficient κ (i.e. imaginary part of complex refractive index $n^* = n + j\kappa$) and hence the optical penetration depth $\zeta = \lambda/4\pi\kappa$ at laser wavelength λ.

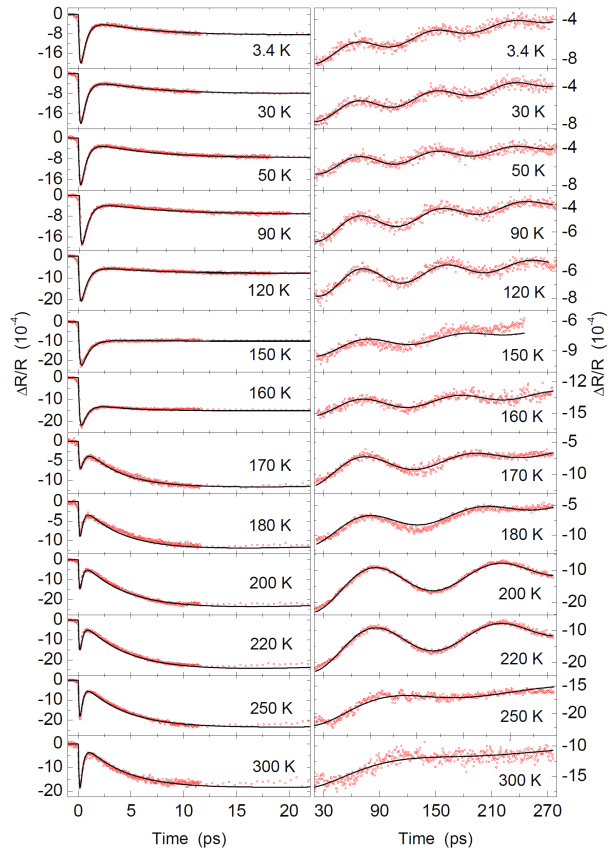

Fig. 1. (Color online) Transient differential reflection spectra from CaFe$_2$As$_2$ as a function of sample temperature taken at a pump fluence of ~85 μJ/cm$^2$. Solid lines are fits using Eq. 1.

*3.1 Quasiparticle dynamics*

First we discuss temperature-dependent dynamics of the exponentially decaying contributions in the differential reflectivity signals attributed to the quasiparticle recombination/relaxation. The results from the time-domain fitting (solid lines in Fig. 1) are presented in Fig. 3, clearly showing presence of one fast component ($\tau_1$) with negative amplitude ($-A_1$), one slow component ($\tau_2$) with positive amplitude ($A_2$) and a much slower component ($\tau_3$) with negative amplitude ($-A_3$), all showing large variations around $T_{SDW}$. As T approaches $T_{SDW}$ from below, slowing down of the relaxation is clear from the increase in $\tau_1$ and $\tau_3$ (Figs. 3b and 3f) whereas decrease of $\tau_2$ at $T_{SDW}$ is evident from Fig. 3d. Such large changes in the temperature-dependence of the three-component QP relaxation amplitudes and time-constants, at first instance, may possibly give a clue to the opening of multiple gaps at $T_{SDW}$ or an associated gap being anisotropic. Indeed, multiple charge gaps in the SDW phase of parent iron pnictides have been inferred from recent infrared spectroscopic[3-5] and ARPES[6] measurements. Theoretically, from the band structure calculations, it is known that the Fermi surface consists of two electron-like pockets at the M-point and three hole-like pockets at the Γ-point of the Brillouin zone.[2,5] Below $T_{SDW}$, opening of a gap in these pockets can give rise to the experimental observations.

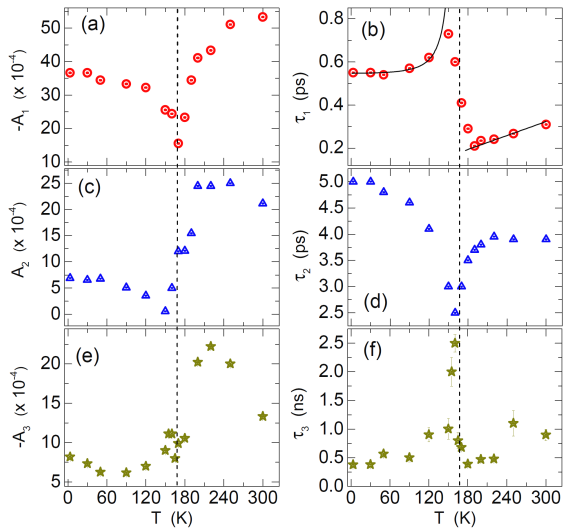

Fig. 3. (Color online) Temperature-dependence of amplitudes $A_k$ and decay times $\tau_k$ of three-component electronic relaxation obtained by fitting the experimental data in Fig. 1 with Eq. 1. The solid lines in (b) are theoretical fits as discussed in the text. The dotted vertical line marks the nominal SDW transition temperature.

Our results for the temperature-dependence of the fast component ($A_1,\tau_1$) in the SDW state, shown in Figs. 3(a,b) are similar to earlier studies on parent iron pnictides[12-14] and can be fitted well within the phonon bottleneck description of QP recombination across a temperature-dependent charge-gap Δ(T) as discussed in the following paragraphs. Moreover, we note another interesting aspect about the QP dynamics from our results in Fig. 3. The amplitudes of all the three relaxation components are unusually higher in the normal state ($T > T_{SDW}$) than in the SDW state ($T < T_{SDW}$). Though the drastic decrease in the amplitudes $A_{1,2}$ while approaching $T_{SDW}$ from below can be attributed to closing of charge gap (gaps) the larger magnitudes of the amplitudes in the normal state of $CaFe_2As_2$ is unusual with respect to the earlier studies on parent iron pnictides[12-14] where the amplitude corresponding to the order parameter was much smaller in the normal state.

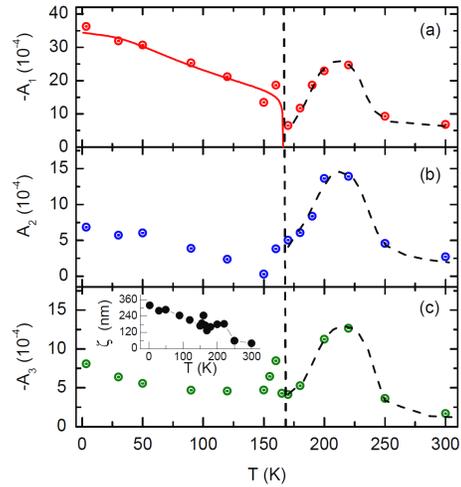

Fig. 4. (Color online) Amplitudes of the three electronic relaxation components ($A_{1,2,3}$) renormalized by the temperature dependent penetration depth ζ(T)/ζ(3.4) at 790 nm as shown in the inset of (c). The solid line in (a) is theoretical fit using Eq. (2) and the dashed lines in (a-c) are drawn as guide to the eyes.

Our results on the coherent longitudinal acoustic phonon mode (Section 3.2) analyzed using propagating strain pulse model[17] give the temperature-dependent optical penetration depth ζ as shown in the inset of Fig. 4c. It is seen that ζ is ~42 nm at the room temperature and increases to large values at low temperatures. Therefore, due to the changes in the optical penetration depth as a function of temperature, the amplitudes $A_{1,2,3}$ of the three QP relaxation components have to be normalized with respect to the absorbed laser energy per unit volume in the sample. The renormalized values of the amplitudes $A_{1,2,3}$ with respect to ζ(T)/ζ(3.4K) are plotted in Figs. 4a-c. It can be seen from Figs. 4a-c that the renormalization does not change the behavior of the amplitudes at temperatures below $T_{SDW}$. However, at higher temperatures, the amplitudes show a peak feature at ~220 K (dashed lines in Figs. 4a-c have been drawn as guide to the eyes). This feature is strikingly similar to that in the temperature dependence of the acoustic phonon amplitude B in Fig. 7a. It appears that presence of spin fluctuations above $T_{SDW}$ can be responsible for such an observation which strongly suggests existence of a characteristic temperature scale of ~220 K in $CaFe_2As_2$. The question to be asked is whether such a feature is related to strong spin-phonon coupling just above $T_{SDW}$ in the normal state of $CaFe_2As_2$ or there exists a precursor pseudo state in the normal phase. In fact a peak in the coherent acoustic phonon amplitude B at a temperature higher than $T_{SDW}$ (Fig. 7a) can be attributed to strong magnetoelastic coupling. Interestingly, in a recent report, Kim et al.,[20] showed that intense femtosecond laser pulses can induce transient SDW order parameter in the normal state of $BaFe_2As_2$, which again was explained in terms of strong spin-phonon coupling between spin fluctuations in the normal state and the optical phonons.

Generally, two types of processes are considered by which quasiparticle creation across a charge gap Δ can take place, either by heating the sample or by optical excitation. Following both the processes, we have an excess density of quasiparticles and our ultrafast optical reflectivity measurements are providing one method to quantify the excess density of quasiparticles. At constant optical excitation fluence, the change in the temperature-dependence of the photoinduced reflectivity provides the temperature dependence of the quasiparticle density and hence the recovery of the gap parameter Δ on reducing the temperature. Dynamics of the quasiparticle recombination across charge gap Δ is governed by phonon creation and reabsorption as described in the Rothwarf-Taylor (RT) model.[15] Here, we use the phonon bottleneck model derived by Kabanov et al.,[16] for the temperature dependence of QP dynamics in charge-gapped systems considering that the photoexcited quasiparticle density at a fixed pump-fluence Φ is substantially less so that they make a small perturbation to the distribution functions of the normal state quasiparticle density. The amplitude (A) of photoinduced reflectivity signal is given by[16]

$$A \propto \frac{\Phi/(\Delta(T)+k_B T)}{1+m\sqrt{k_B T/\Delta(T)}\exp[-\Delta(T)/k_B T]} \quad (2)$$

where m is a material-dependent constant. Assuming a BCS-like form $\Delta(T) = 2\Delta_0\{1-(T/T_{SDW})^2\}$ where $2\Delta_0$ is the zero-temperature value of the SDW gap parameter, Eq. (2) describes an increase in the photoexcited quasiparticle density due to the decreasing gap value and corresponding enhanced phonon emission during the initial relaxation. As the gap closes at $T_{SDW}$, more low-energy phonons become available for reabsorption and the quasiparticle recombination mechanism becomes less and less efficient resulting in a quasi-divergence in the relaxation time given as[16]

$$\tau \propto \frac{\ln(g+\exp[-\Delta(T)/k_B T])}{\Delta(T)^2}, \quad (3)$$

where g is a fitting parameter. Equation (2) is fitted to $A_1(T)$ in Fig. 4(a) for $T < T_{SDW}$ giving $2\Delta_0/k_B T_{SDW}$ = 1.6±0.2. The solid line in Fig. 3(b) for $T < T_{SDW}$ is a fit to Eq. (3) with same value of $2\Delta_0/k_B T_{SDW}$. The value of $2\Delta_0/k_B T_{SDW}$ being less than the weak coupling limit of the BCS theory implies that there can be more than one gap opening at the SDW transition because in the phonon bottleneck model, the hot QP relaxation is governed by the smallest energy gap in the system due to strong interband scattering.[14] The absence of a divergence in $\tau_1$ as T → 0 unlike in heavy Fermionic SDW compound $UNiGa_5$,[21] indicates presence of ungapped Fermi surfaces at T < $T_{SDW}$.[14]

We may note that Eqs. (2) and (3) give the QP dynamics for weak perturbation,[16] i.e., the photoinduced carrier density $n_s$ being smaller than the thermally generated carrier density $n_T$. In our case, we can estimate the photoinduced carrier density using $n_s = \Phi\alpha/h\nu_{probe}$ where Φ is the fluence and α = 1/ζ. From our present experimental results on the coherent acoustic phonons we have self-consistently derived a value of ζ ~ 320 nm at the lowest temperature of 3.4 K. Taking Φ = 100 μJ/cm$^2$ and ζ ~ 320 nm, unit cell volume[22] = 170 Å$^3$ we obtain $n_s$ ~ 1.5x10$^{-3}$ /unit cell. Typical quasiparticle concentration in $CaFe_2As_2$ can be estimated using the relation[16] $n_0$ = 2N(0)Δ where N(0) is the density of states at the Fermi energy. For $CaFe_2As_2$, using N(0) ~ 5 /eV/unit cell[23] and Δ = 1.6$k_B T_{SDW}$ ~ 20 meV, we obtain $n_0$ ~ 0.20 /unit cell which is much larger than $n_s$, thereby justifying the use of the phonon-bottleneck model to analyze our results for $A_1$, $\tau_1$ while extracting the value of the band gap $2\Delta_0$.

The slow component $(A_2,\tau_2)$ in our studies (Figs. 3c and 3d) is unique and new as compared to the prior studies on parent iron pnictides.[12-14] Here the amplitude $A_2$ has similar temperature-dependence as that of the fast one $(A_1)$ but the decay time $\tau_2$ shows a converging trend as T → $T_{SDW}$ rather than the diverging behavior of $\tau_1$. Similar competing behavior of the fast and slow relaxation processes was seen in charge density wave systems[24] where the fast component with time constant τ ~ 0.5 ps diverging at $T_{CDW}$ was attributed to quasiparticle recombination dynamics across the CDW gap whereas the slow relaxation with time constant τ ~ 7 ps showed a large decrease while approaching $T_{CDW}$ from below assigned to the second stage of the CDW recovery. In our case, the observation of competing behavior of the fast $(\tau_1)$ and slow $(\tau_2)$ relaxation processes may be an indication of two sets of quasiparticles, either on different Fermi sheets around the Γ and/or M points or separately on two different Fermi sheets around these points in the Brillouin zone, competing for the density of states in the excited state. However, more understanding has yet to emerge.

The slower component $(A_3,\tau_3)$ can arise due to heat diffusion out of the photoexcited volume as suggested in

Sr-122 and Sm-1111 iron pnictides[13,14] as well as from charged quasiparticle interaction with acoustic phonons and spin degrees of freedom.[12]

So far we focused our discussion on the temperature-dependent QP dynamics in the SDW state where the sharp changes in the amplitudes and corresponding time-constants of all three relaxation components at $T_{SDW}$ manifest opening of charge gaps at that temperature. In the normal paramagnetic-metallic state ($T > T_{SDW}$), the temperature-dependent increase in the decay-time $\tau_1$ (Fig. 3b) can be used to estimate the electron-phonon coupling constant $\lambda$ using the relation[14] $\tau = 2\pi k_B T / 3\hbar \lambda \langle \hbar^2 \omega^2 \rangle$ where $\lambda \langle \hbar^2 \omega^2 \rangle$ is the second moment of the Eliashberg function. The linear fit (continuous line in Fig. 3b) gives $\lambda \langle \hbar^2 \omega^2 \rangle \sim 80$ meV$^2$. Using the Eliashberg spectral function for the Ca-122 system[25] we obtain $\langle (\hbar \omega)^2 \rangle \sim 610$ meV$^2$. Therefore, we get the electron-phonon coupling constant $\lambda \sim 0.13$, a case of weak electron-phonon coupling. This value of $\lambda$ is similar to that obtained for other parent iron pnictides.[13,14]

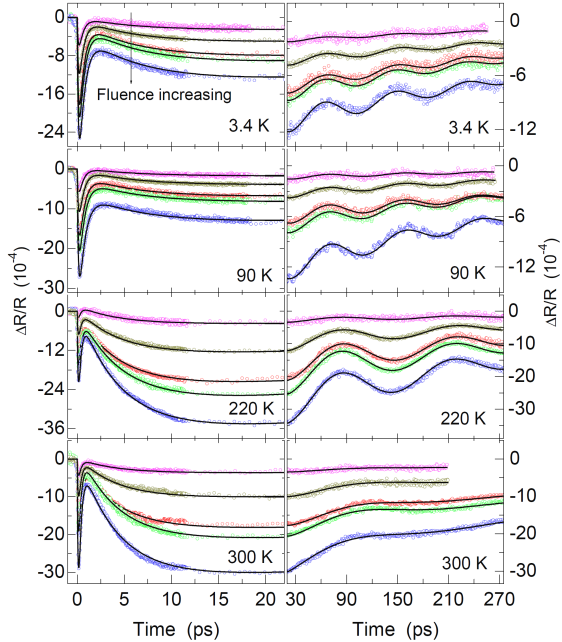

Fig. 5. (Color online) Transient differential reflectivity as a function of the pump-fluence at four representative temperatures, two in the SDW and other two temperatures in the normal state of the sample. In each panel the data are plotted for five different values of the pump-fluence (20, 50, 85, 110, 175 μJ/cm$^2$) increasing in the direction of the black arrow.

Time resolved differential reflectivity data at various pump-fluences taken at four representative temperatures of 3.4 K, 90 K, 220 K and 300 K are presented in Fig. 5. The fluence-dependences of the relaxation parameters obtained by fitting the data (fits are shown by solid lines in Fig. 5) using Eq. (1) are plotted in Fig. 6 in both the $T < T_{SDW}$ and $T > T_{SDW}$ regions. We note that the amplitude renormalization with respect to the temperature-dependent penetration depth as discussed before will not change the behavior of the fluence-dependence hence the amplitudes $A_{1,2,3}$ in Fig. 6 have been plotted as obtained from the time-domain fitting. In the normal state ($T > T_{SDW}$), the amplitudes increase almost linearly with the increasing fluence while the decay times remain fluence-independent. However, for $T < T_{SDW}$, saturation can be seen for all the three amplitudes (Figs. 6a, 6c, and 6e). Also, the decay time $\tau_1$ increases with the fluence (Fig. 6b). In the SDW state, linear increase in the amplitudes at low fluences and $\Phi^{1/2}$ dependence beyond a threshold indicate the gapped nature of the associated charge states as per the RT model.[26] However, the increasing behavior of $\tau_1$ (Fig. 6b) is inconsistent with the RT model which rather predicts a linear increase in the quasiparticle relaxation rates, i.e., $1/\tau \propto \Phi$.[26] In the prior studies on the parent compounds, the time constants associated with the carrier recombination in the SDW state were found to be fluence-independent[13,14] whereas, amplitude-saturation similar to ours was observed. The inconsistency with the RT model is not obvious from the present understanding of our temperature and fluence-dependent quasiparticle relaxation dynamics and needs to be explored further.

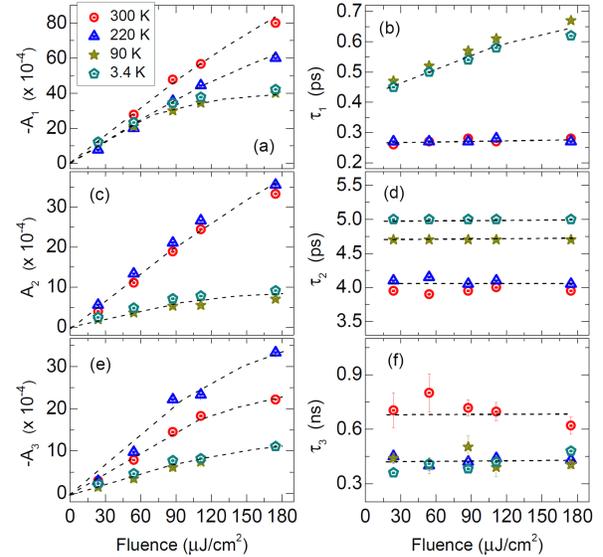

Fig. 6. (Color online) Pump-fluence dependence of the amplitudes $A_k$ and decay times $\tau_k$ of the three electronic relaxation components shown for temperatures from both the SDW and normal states of CaFe$_2$As$_2$.

At this stage we must highlight the important outcome from our study on the QP dynamics in parent CaFe$_2$As$_2$ as compared to the recent ultrafast optical studies on other parent iron pnictides. First of all, use of low repetition rate laser pulses at 1 kHz avoids the problem of accumulation of laser heating and is in contrast to the previous reports. In all the prior ultrafast studies on similar systems, researchers have used high repetition rate laser pulses where one needs to focus the laser beams tightly on to the sample to achieve fluences of a few tens of μJ/cm$^2$. In the present study, we observe three QP relaxation components showing dramatic changes at $T_{SDW}$. Also, the relaxation component at intermediate time-scales of ~5 ps is unique to the CaFe$_2$As$_2$ system where the quasiparticle relaxation becomes faster while approaching $T_{SDW}$. On the other hand, for Ba-122 or Sr-122 parent compounds[12,14] a single electronic relaxation component with amplitude following BCS-like temperature dependence having $2\Delta_0 \sim 3.5k_BT_{SDW}$ has been reported and only in the doped compounds of these systems, either two or three relaxation components were observed[27-29] in comparison with three-component QP relaxation dynamics in CaFe$_2$As$_2$. These differences suggest that the nature of SDW gaps in Ca-122 systems can be different from other 122 systems. In fact, the temperature dependent resistivity in CaFe$_2$As$_2$ is interestingly different from that in other parent iron pnictides; namely that a steep increase is seen in the resistivity[7,8] at $T_{SDW}$ rather than a decrease as shown by Ba-122 and other types of pnictides.[7] Whether such a difference in the resistivity has any implications in the optical response from the two compounds, an understanding has yet to emerge from further studies.

### 3.2 Coherent longitudinal acoustic phonons

Temperature and fluence-dependences of amplitude B, frequency $\nu_p$, dephasing time $\tau_p$ and initial phase $\phi$ of the coherent oscillations in the transient reflectivity data (Figs. 1 and 5) obtained from the time domain fitting (Eq. 1) are presented in Fig. 7(a-d) and Fig. 7(e-h), respectively, where dashed lines have been drawn as guide to the eyes. It is noteworthy to note that all the four parameters are highly temperature dependent across the SDW transition (Fig. 7a-d) however they remain nearly fluence-independent (Fig. 7e-h) except the amplitude B which shows saturation with increasing fluence beyond ~120 μJ/cm$^2$. In the vicinity of $T_{SDW}$, i.e., at T ~ 220 K, the amplitude shows maximum coupling with the incident laser energy.

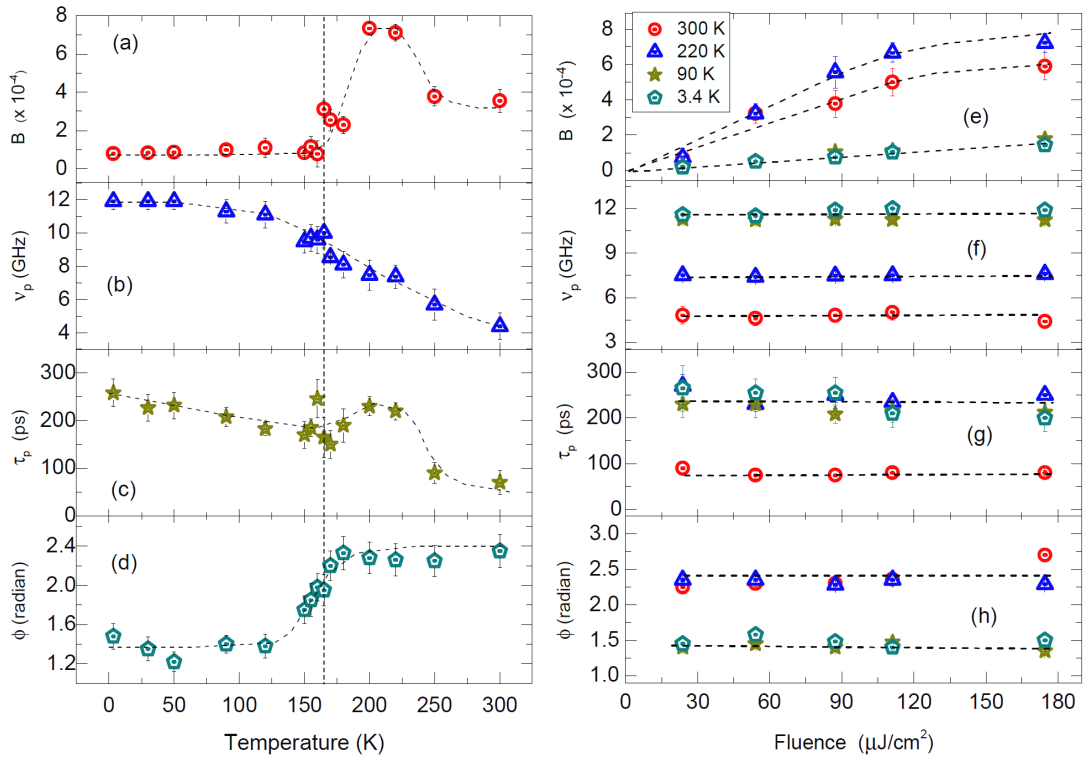

Fig. 7. (Color online) Temperature and pump-fluence dependence of the amplitude B, damping time $\tau_p$, frequency $\nu_p$ and initial phase $\phi$ of the coherent longitudinal acoustic phonon mode generated in CaFe$_2$As$_2$. The solid lines are guide to the eyes.

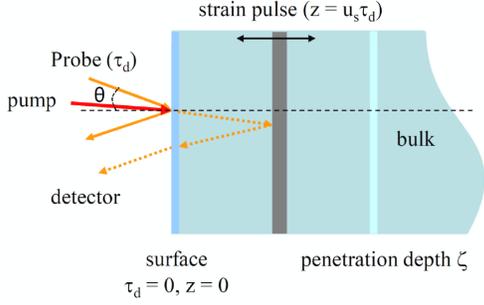

Fig. 8. (Color online) Schematic of the launching of a coherent strain wave in the direction normal to the surface in a propagating strain-pulse model. A strong pump-pulse creates a thermal strain at the surface (z = 0) which propagates into the medium with an average sound velocity $u_s$. Interference between the time-delayed ($\tau_d$) probe pulses partially reflected from the interfaces defined by the front surface and the stain-pulse causes oscillations in the transient reflectivity signal.

Generally, hot carriers generated by an ultrafast pump-pulse excitation can relax their energy via emission of phonons which cause oscillatory changes in the transient reflectivity signal. Coherent oscillations can also be induced by electronic and/or thermal stress at the photoexcited sample surface leading to strain pulse propagation into the sample with sound velocity $u_s$. The generation process of longitudinal acoustic phonons by strain-pulse propagation in the iron pnictide crystals is depicted in Fig. 8. The pump-pulse energy deposited at the surface of the crystal gives rise to transient electron and phonon-temperature rise which then set up a transient stress via thermoelastic and/or electronic stress.[17,30,31] This stress is released by launching a strain wave in the form of coherent longitudinal acoustic (LA) wave propagating with sound velocity $u_s$ into the crystal in the direction normal to the surface. The interference between the probe beams partially reflected from the interfaces defined by the crystal surface and the propagating strain pulse within the optical probe penetration depth is responsible for the detectable coherent oscillations (right panels of Figs. 1 and 5) in the reflectivity signal (for angle of incidence $\theta = 0$)[30]

$$\Delta R/R \propto k_{pr} \cos(2nk_{pr}u_s t + \phi), \quad (4)$$

$k_{pr}$ being the probe wave-propagation vector, n the refractive index and $\phi$ the phonon phase. Therefore, the frequency of oscillations at a fixed probe angle of incidence $\theta$ is directly related to the sound velocity $u_s$ and the probe wavelength $\lambda_{pr}$ through the relation

$$\nu_p = 2nu_s \cos\theta / \lambda_{pr}. \quad (5)$$

In the SDW phase the magnitude of the phonon amplitude B is nearly constant and much smaller possibly due to smaller thermal expansivity as compared to that in the normal state (Fig. 7a). Around $T_{SDW}$, B starts increasing and attains a peak feature at a temperature of ~220 K. This peak feature in the phonon-amplitude B at a temperature higher than $T_{SDW}$ is similar to that was discussed before for the electronic relaxation components (Fig. 4) and is attributed to precursors of spin fluctuations in the normal paramagnetic phase.[32,33] Large magneto-elastic coupling between the short range spin fluctuations and lattice distortions can occur near the onset of SDW phase in the iron pnictide lattice,[34] an effect similar to that was observed in multiferroic manganites above the magnetic transition temperature.[18,19] It is noted from Fig. 7b that the phonon-frequency $\nu_p$ is nearly constant in the deep SDW state but it starts decreasing as $T_{SDW}$ is approached from below. The corresponding change in the phonon-phase $\phi$ by $\sim\pi/2$ (Fig. 7d) is a consequence of the structural transition.

The acoustic phonon-dephasing time $\tau_p$ also displays substantial temperature-dependence across $T_{SDW}$ (Fig. 7c). In the strain pulse model, the dephasing of the oscillations can arise due to a combination of intrinsic phonon life time $\tau_{ph}$ and another contribution $\tau_\alpha$ related to the probe absorption ($\alpha$) within the optical penetration-depth $\zeta$ in the crystal.[31] The phonon life-time corresponds to a time when the strain-pulse can propagate into the crystal within its characteristic length before complete energy transfer takes place from the coherent mode to a distribution of incoherent modes. Similar observations of temperature-dependent coherent longitudinal acoustic vibrations with frequency 3 to 4.2 GHz were reported before in thin films (thickness ~450 nm) of superconducting $FeSe_{1-\delta}$ iron pnictide[35] where the time-period of oscillations were the round trip time of the coherent phonon propagation across the film thickness. For bulk samples, as is the case here, the strain pulse can propagate through the entire thickness of the sample and hence the observed phonon-dephasing $\tau_p$ in Fig. 7(c) should be limited by $\tau_\alpha$ due to finite probe penetration depth. In fact, this provides an indirect way of measuring optical penetration-depth $\zeta = u_s\tau_p$ where $u_s$ is the sound velocity and $\tau_p$ is identified as $\tau_\alpha$. We have to point out categorically that there is no report of the experimental value of the optical penetration depth at 790 nm at any temperature for Ca-122 systems in the literature. Sometimes the observed temperature dependences of the phonon dephasing time and phase can be attributed to the temperature dependent changes of the refractive index n and extinction coefficient $\kappa$ as observed by Lim et al.,[36] for $LuMnO_3$ crystal where the probe wavelength is in proximity of the d-d optical transition peak that causes the refractive index and the extinction coefficient vary strongly with temperature. Based on the strain pulse propagation in combination with reflectivity value R, our experiments help to self-consistently estimate the temperature dependence of the refractive index n and the extinction coefficient $\kappa$ (and hence optical penetration depth $\zeta$) at 790 nm as described below.

We performed linear reflectivity measurements on our $CaFe_2As_2$ single crystal by focusing the femtosecond laser beam (low average power) on the flat portion of the sample and collecting all the reflected light for near-normal incidence. The measured value is R = 0.34 which is constant as a function of temperature within an accuracy of ±5%. As mentioned before, to be consistent with the strain pulse propagation model, the observed phonon dephasing at temperature T should correspond to the optical penetration depth ζ via the relation

$$\zeta(T) = u_s(T)\tau_p(T), \quad (6)$$

where the sound velocity $u_s(T)$ is related to $\nu_p(T)$ and the refractive index $n(T)$ via Eq.(5). Since $\zeta = \lambda_{pr}/4\pi\kappa$, by incorporating Eq.(5) for normal incidence $(\theta = 0)$ into Eq.(6), this relation simply leads to

$$\frac{n(T)}{k(T)} = 2\pi\nu_p(T)\tau_p(T) \quad (7)$$

For normal incidence, the Fresnel's equation for reflectivity is given as

$$R(T) = \frac{(n-1)^2 + \kappa^2}{(n+1)^2 + \kappa^2} \quad (8)$$

Using constant value of R = 0.34±0.02 during 3.4 K to 300 K, we have estimated κ and n as a function of temperature T from Eqs. (7) and (8), shown in Figs. 9(a) and 9(b). The temperature dependent optical penetration depth ζ can be directly calculated from κ(T) as shown in the inset of Fig. 4(c). It is noteworthy to see that the penetration depth changes from ~320 nm at 3.4 K to ~42 nm at 300 K leading to important consequences on the renormalization of the amplitudes ($A_{1,2,3}$) as shown in Figs. 4(a-c). Previous results on the superconducting $FeSe_{1-\delta}$ films[35] also indicated that the low temperature optical penetration depth (absorption) is atleast an order higher (lower) than the value at room temperature. Importantly, without any assumption, we have estimated the temperature dependence of n and κ from the experimental results for the frequency $\nu_p$ and dephasing time $\tau_p$ of the acoustic phonon mode. We hope that our results will motivate detailed temperature dependent ellipsometry measurements of the optical constants of Ca-122 systems in future.

By knowing the temperature dependence of phonon frequency $\nu_p$ and refractive index n, as discussed above, the temperature dependence of the sound velocity $u_s$ can be calculated using Eq. (5) as shown in Fig. 9(c). The elastic modulus is obtained by using $Y = \rho.u_s^2$, where ρ is the density of the crystal. The temperature-dependence of the density ρ(T) of the crystal has been measured by x-ray diffraction which shows that unit cell volume of the $CaFe_2As_2$ lattice increases sharply[22] at $T_{SDW}$. We have used the mass of a unit cell $M_{unit-cell}$ ~ 96.98 x$10^{-23}$ g and the temperature-dependent unit cell volume of $CaFe_2As_2$ from ref. [22] in estimating the density ρ (shown in the inset of Fig. 9d) and hence the elastic modulus Y of the crystal as presented in Fig. 9(d). The temperature-dependence of the elastic modulus Y as shown in Fig. 9(d) is different than that measured using ultrasound measurements at MHz frequencies.[37,38] The latter shows that the elastic constants decrease as T → $T_{SDW}$ from either side of $T_{SDW}$. This difference can arise from the difference in the coupling of sound waves of two frequencies with the slow relaxation processes in the system leading to dissipation and elastic dispersion.[39] In our case, the system relaxation time is as large as 500 ps (~$\tau_3$), which means that the observed acoustic waves in present experiments (ω ≥ 1/$\tau_3$) are in the zero-sound region ($c_0$). On the other hand, the ordinary adiabatic (or first-sound) velocity ($c_1 < c_0$) is valid for ω << 1/$\tau_3$. Therefore, the difference in the elastic properties measured using the present technique of ultrafast ultrasonics (GHz and above) and ultrasound spectroscopy (MHz and below) will depend on the difference between $c_1$ and $c_0$ and their temperature-dependences.

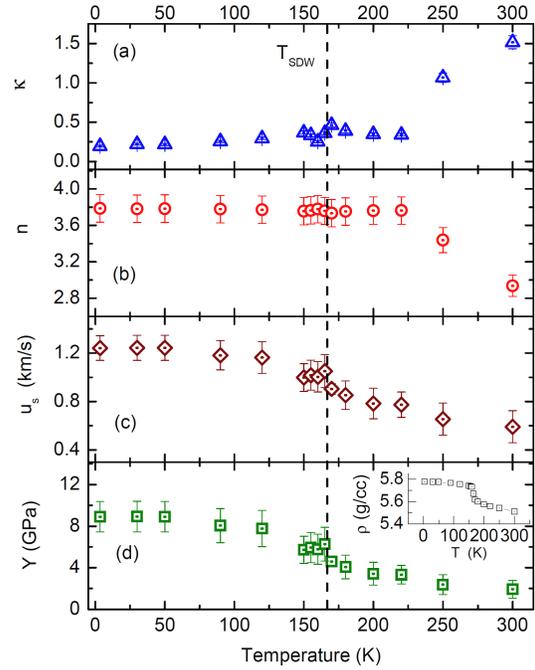

Fig. 9. (Color online) Temperature-dependent (a) extinction coefficient κ, (b) refractive index n, (c) longitudinal sound velocity $u_s$, and (d) elastic modulus Y derived from the temperature dependence of the acoustic phonon parameters as described in the text. The inset in (d) is the density ρ of the crystal taken from ref. [22] in combination with mass of unit cell $M_{unit-cell}$ ~ 96.98 x$10^{-23}$ g that has been used to estimate the elastic modulus Y.

## 4. Conclusions

Three-component quasiparticle relaxation is observed in $CaFe_2As_2$ iron pnictide single crystal on both sides of the spin density wave phase transition temperature. The amplitudes and time constants of all three relaxation components show large variations around $T_{SDW}$. By analyzing the dynamics of the fast component using well known phonon bottleneck model, we have estimated the zero-temperature value of the SDW charge gap to be ~ $(1.6\pm0.2)k_B T_{SDW}$. Further, from the normal state temperature-dependence of the fast relaxation time $\tau_1$, electron-phonon coupling constant is estimated to be ~0.13. Our transient reflectivity data have clearly shown coherent longitudinal acoustic phonon oscillations over the entire temperature range that helped us in determining the crystal elastic behavior in both the normal as well as spin density wave phase.

**Acknowledgement**

AKS and SK acknowledge Department of Science and Technology, India for financial assistance.

Electronic address: §*sunilvdabral@gmail.com*,
*asood@physics.iisc.ernet.in